\documentclass[prd,aps,preprintnumbers,nofootinbib]{revtex4}
\usepackage{epsfig}
\usepackage{amsmath}
\usepackage{hyperref}

\def\bea {\begin{eqnarray}}
\def\eea {\end{eqnarray}}
\def\nn{\nonumber}

\def \dq {\delta q}
\def \lrg {{\stackrel{\leftrightarrow}{\gamma}}}

\begin{document}

\preprint{CERN-PH-TH/2012-149, YITP-SB-12-29}

\renewcommand{\thefigure}{\arabic{figure}}

\title{Final state interactions in single- and multi-particle inclusive cross sections\\ for hadronic collisions}

\author{Alexander Mitov}
\affiliation{Theory Division, CERN, CH-1211 Geneva 23, Switzerland}
\author{George Sterman}
\affiliation{C.N.\ Yang Institute for Theoretical Physics, Stony
Brook University, Stony Brook, New York 11794--3840, USA}

\date{\today}

\begin{abstract}
We study the role of low momentum transfer (soft) interactions between high-transverse momentum heavy particles and beam remnants (spectators) in hadronic collisions.   Such final-state interactions are power suppressed for single-particle inclusive cross sections whenever that particle is accompanied by a recoiling high-$p_T$ partner whose momentum is not fixed.  An example is the single-top inclusive cross section in top pair production.   Final-state soft interactions in multi-particle inclusive cross sections, including transverse momentum distributions, however, produce leading power corrections in the absence of hard recoiling radiation. Nonperturbative corrections due to scattering from spectators are generically suppressed by powers of $\Lambda/p'_T$, where $\Lambda$ is a hadronic scale, and $p'_T$ is the {\it largest} transverse momentum of radiation recoiling against the particles whose momenta are observed.
\end{abstract}
\maketitle

\section{Introduction}

At hadron colliders, final states with high transverse momentum and massive strongly-interacting particles play an important role in the search for physics beyond the standard model, but present special challenges to theory.    Our best theoretical predictions apply when cross sections can be computed using conventional collinear factorization \cite{Collins:1989gx}, with corrections suppressed by powers of a hard scale.    Such factorization has been shown for the inclusive production of electroweak bosons \cite{Collins:1989gx,Bodwin:1984hc,Aybat:2008ct,Collinsbook} and single-particle inclusive (1PI) cross sections in hadronic collisions \cite{Nayak:2005rt}.   On the other hand, how far specific observables may be generalized while retaining factorizability with small corrections is not fully understood \cite{Catani:2011st}.  
While proofs of factorization require a full treatment of soft and collinear radiation, one of the basic ingredients is the cancellation of the final-state interactions of the observed particles.     In this paper, we will concentrate on corrections due to final state interactions in single- and multi-particle inclusive cross sections in hadronic collisions.

We present below an analysis based on light cone ordered perturbation theory of interactions between hard, final-state partons and spectator partons (remnants) from initial state hadrons.     We observe that the cancellation of final-state interactions involving one or more observed final-state particles requires that the cross section be insensitive to changes in the recoil momentum of unobserved particles.
This analysis suggests an estimate of the residuals of the cancellation, and under what circumstances we  can anticipate significant and perhaps measurable effects.   We will see that the cancellation of final state interactions (FSI) requires that untagged particles produced in the hard collision carry sufficient momentum transverse to the beam axis to absorb the recoil of momenta transferred by these interactions to the observed particles.  For top or antitop 1PI cross sections in top pair production, the untagged partner plays this role.   
Uncanceled FSI effects are suppressed by at least a single power of $m_t$, consistent with a recent estimate of nonperturbative string-breaking effects in Ref.\ \cite{Rosner:2012pi}.  
Indeed, our analysis is inspired in part by the apparent mismatch between observed top pair asymmetries \cite{Orbaker:2011hz} and Standard Model predictions based on factorized cross sections \cite{Kidonakis:2011zn,Almeida:2008ug,Ahrens:2011uf}.   We observe, however, that this mismatch is primarily in terms of normalization, and that the roughly linear dependence  of the asymmetry on both pair invariant mass and rapidity difference is shared by resummed QCD predictions \cite{Almeida:2008ug} and the data as reported, for example, in \cite{Mietlicki:2012uh}.   It has also been suggested that differences in normalization could be due to the choice of renormalization and factorization scales in the predictions \cite{Brodsky:2012ik}.   Final state interactions seem therefore unlikely candidates for an explanation of top quark asymmetry measurements based on single-particle inclusive cross sections.   The more general question of when such corrections may be important, however, is of independent interest.

We will argue 
that for double-particle inclusive cross sections, involving for example both members of a top pair, the cancellation of FSI requires summation over states with additional high transverse momentum radiation.   We will see that it is the transverse momenta of this radiation, rather than the mass or transverse momenta of the top pair, that controls the scale of power corrections to the factorized forms of cross sections.    At low enough values of recoiling transverse momentum, collinear factorization is no longer possible.

The need to generalize collinear factorization whenever recoiling momentum is small compared to the scale of the hard scattering is familiar from transverse momentum distributions in electroweak boson production at hadron colliders \cite{Dokshitzer:1978yd,Collins:1984kg}. At low values of the $Z$ or $H$ transverse momentum distribution collinear factorization must be replaced by factorization in terms of transverse momentum-dependent parton distributions (TMDs), with explicit demonstrations given in Refs.\  \cite{Laenen:2000ij,Collinsbook}.    TMDs are closely related to the  beam function formalism developed in \cite{Stewart:2009yx}.      Recently, however, it was shown by Collins and Qiu \cite{Collins:2007nk} and Rogers and Mulders \cite{Rogers:2010dm} that TMD factorization is not possible for the double inclusive cross sections of strongly interacting particles when scattering from spectators is taken into account.    References \cite{Collins:2007nk} and \cite{Rogers:2010dm} proceeded by a careful analysis of low-order diagrams.   Here we provide an all-orders analysis of the conditions necessary specifically for leading power collinear factorization in single- and multi-particle inclusive corrections.  

We will frame our discussion in the context of top-pair production at hadron colliders, for its intrinsic interest, treating final state interactions to all orders in perturbation theory.  Our considerations are simple but realistic, however, and the conclusions have much wider applicability. A related observable is the same sign dimuon asymmetry \cite{Mitov:2011us} measured at the Tevatron \cite{Abazov:2010hv}.

In the following section, we review the classification of initial and final states according to light-cone ordered perturbation theory (LCOPT), to
illustrate the origin of long-distance effects and to study final state and other long distance corrections in a general context.    The cancellation of final state interactions in single-particle inclusive cross sections is discussed in Sec.\ \ref{sec:NLO}, first at lowest order and then to all orders.    We find a pattern for the cancellation of final state interactions that illustrates the role of recoiling final state partons, and which leads us to conclude that  corrections to leading-power factorized cross sections are suppressed by the top quark mass in this case.  In contrast, we show in Sec.\ \ref{sec:extensions} that the cancellation of final-state interactions for two-particle inclusive cross sections depends on the presence of additional high $p_T$-radiation, which also sets the scale for power-suppressed corrections.  We conclude with a discussion of our results, their possible extensions and phenomenological implications in Sec.\ \ref{sec:discussion}.

\section{Top pair final states in light cone ordered perturbation theory}
\label{sec:lcopt}

As noted in the introduction, we will consider massive-pair inclusive cross sections of the form,
\bea
H_A(p_A)+H_B(p_B) \rightarrow t(p_t)+\bar t(p_{\bar t})+X\, ,
\label{eq:process}
\eea
 with top production in mind, although our
reasoning is more general.    To be specific, when we
consider a single-particle inclusive cross section, we fix the top quark momentum, and
integrate over the anti-top quark momentum.
The top quark mass, $m_t$ already provides a hard scale, and to avoid multiple scales we imagine that
the transverse momentum of each of the quarks in the pair is of order $m_t$.
For the single-particle inclusive cross section, we then have a factorized expression \cite{Nayak:2005rt}
\begin{equation}
\frac{d\sigma}{d^3p_t} = \sum_{ab} f_{a/A} \otimes f_{b/B} \otimes \hat \sigma_{ab\rightarrow t+X}^{\rm part}(p_t) + {\cal O}(\Lambda/m_t) \, ,
\label{eq:FTh}
\end{equation}
where for this discussion we consider only top pair, not single-top, processes.
The factorized cross section is a convolution in momentum fractions of a perturbatively calculable hard function, $\hat \sigma^{\rm part}_{ab\rightarrow t+X}$ and nonperturbative but process-independent parton distribution functions $f_{c/C}$, for each parton $c$ in hadron $H_C$.  Corrections are present, but  are suppressed by the hard scale $m_t$, as we shall verify below.
The arguments for factorization in single-particle inclusive cross sections of this sort were 
assembled
in Ref.\ \cite{Nayak:2005rt} for hadrons produced in fragmentation.     The case of the top quark 
is actually simpler, because the top quark pair is produced at short distances,
and there is no need of a nonperturbative fragmentation function.   
Nevertheless, the top quark and antiquark do undergo interactions before they decay,
and one of our goals is to show how these effects cancel in (\ref{eq:FTh}).
We will then go on to study multi-particle inclusive cross sections, starting with the two-particle
case where both the top and anti-top momenta are observed,
\begin{equation}
\frac{d\sigma }{d^3p_td^3p_{\bar t}} = \sum_{ab} f_{a/A} \otimes f_{b/B} \otimes \hat \sigma_{ab\rightarrow t+\bar t + X}^{\rm part}(p_t,p_{\bar t}) + C_{2PI} \, ,
\label{eq:FTh2}
\end{equation}
and to estimate the size of corrections, $C_{2PI}$ associated with the final state interactions of the top pair.

\subsection{The notation of LCOPT}

To quantify the effect of final-state interactions we use light-cone ordered perturbation theory (LCOPT) \cite{Chang:1968bh},
relying on much of the same algebraic analysis as applied to jet cross sections in Ref.~\cite{Sterman:1978bj}.
Effectively, in LCOPT the integration over the
the minus (or plus) light-cone components,
\bea
k^\pm = \frac{1}{{\sqrt2}}\left (k^0 \pm k^3\right)
\eea
 is carried out for each line momentum. 
 
 The result of minus integrals is a sum over diagrams with $x^+$-ordered vertices, separating states 
 with lines whose minus momenta are fixed by the mass-shell condition.
 The characteristic feature of LCOPT is that all lines move `forward' in $x^+$, with positive plus momenta only
 \cite{Brodsky:1997de,Langnau:1991iq,Berger:2003zh}.
 For the production of a top quark pair,  final states will be those that include the pair, and by implication,
initial states are those that do not.
To present the resulting diagrams, it is convenient to introduce the notation
\bea
[k]^- &\equiv& \frac{m^2+k_{\perp}^2}{2k^+}\, ,
\label{eq:bracketdef}
\eea
with $m$ the mass and $k$ the momentum.

In LCOPT,  a single covariant diagram ${\cal G}_{ \{ p\} \rightarrow \{q\} }$ with four-dimensional 
loop integrals is rewritten as the sum of diagrams in which all vertices are ordered (in $x^+$), 
and in which plus momentum integrals extend over only that range in which all lines flow forward
(in $x^+$).    Such a  diagram can then be written as a sum over vertex orderings, $T$.   Each ordering prescribes a set
of states $s$,  consisting of lines that appear between two vertices that are neighbors in the ordering.
We represent it as
\bea
{\cal G}_{ \{ p\} \rightarrow \{q\} }
&=&
\sum_{{\rm orderings}\ T}\
\int \prod_{{\rm loops}\,\{l\}} d^2l_{\perp} dl^+\ \prod_{{\rm lines}\,\{k\}}\ \frac{\theta(k^+)}{2k^+}\ \prod_{{\rm states}\,\{s\}\ {\rm in}\ T} \frac{1}{P^- - s\left ( [k]\right)+i\epsilon }\ N\left ( \{p\},\{q\},[k]\right)\, ,
\label{eq:lcopt}
\eea
where $P^-=\sum_a p_a^-$ is the total incoming minus momentum, and where
\bea
s\left ( [k]\right)\
=
\sum_{{\rm lines}\,\{k\}\, \in\, {\rm state}\, s} [k]^-\, ,
\label{mdef}
\eea
is the sum of all the on-shell minus momenta in a specific state, determined as in (\ref{eq:bracketdef}).   
For any given state, the sum may include a subset of incoming lines
$p_a^-$ and/or (with a negative sign) outgoing lines $q_j^-$.   An overall momentum
conservation delta function and other constants have been suppressed.   The factors $\theta(k^+)$ ensure that plus momenta
flow forward, that is, from earlier to later vertices in the ordered amplitude (and the opposite in the complex conjugate). 
The factor $N\left ( \{p\},\{q\},[k]\right)$ 
represents all overall momentum and constant factors.    We shall
assume that $N$ is a polynomial  in loop momenta, in which case it does not affect our reasoning below.

\subsection{Choice of Frame}

To analyze final-state interactions between remnants of the initial state hadrons and the produced pair,
it will be convenient to treat the momenta of the incoming hadrons on an equal footing.    This forces
us to choose a frame in which the 3-direction that defines the light-cone momentum
component $k^-$, over which we will integrate, is not in the direction of either of the incoming hadrons,
$p_A$ and $p_B$ in Eq.\ (\ref{eq:process}).
To be specific, we will choose a center of mass frame
in which these momenta are in the positive and negative 1-direction, 
so that
\bea
\frac{1}{2}\, \sqrt{\frac{S}{2}} &=& p_A^\pm = p_B^\pm\, ,
\nn\\
\frac{\sqrt{S}}{2}&=& p_A^0 = p_B^0\, ,
\nn\\
p_A^1 &=&  p_A^0\, ,
\nn\\
p_B^1 &=& - p_B^0\, .
\label{eq:in-momenta}
\eea
With this choice of frame, we define for any momentum $k$,
\bea
k_{\perp} = (k^1,k^2)\, .
\label{eq:perp-def}
\eea
Thus, the incoming hadrons start with equal and opposite $\perp$-momentum.
When we want to refer to momenta transverse to the beam direction
we will use the notation, $k_T$,
\bea
k_T = (k^2,k^3)\, .
\label{eq:trans-def}
\eea

\subsection{Cut diagram notation, initial and final states}
\label{sec:cut-diagram}

To construct our cross sections for any given out state with momenta $\{q\}$ from an in state with 
momenta $\{p\}=\{p_A,p_B\}$, we take the absolute square
of the sum of covariant diagrams ${\cal G}_{ \{ p\} \rightarrow \{q\} }$, as they contribute to the $S$-matrix.
We will use the term `out state' to refer to a specific 
contribution to the inclusive cross section, to distinguish
it among the class of `final states', which will refer to any state that includes the produced pair of heavy quarks (or other
high-$p_T$ particles).    We are thus treating the tops as perturbatively-produced rather than as part
of the parton distributions, and for definiteness we are neglecting single-top production, although our analysis 
applies as well to this case.

\begin{figure}[h]
  \centering
     \hspace{-1mm} 
   \includegraphics[width=3.5in]{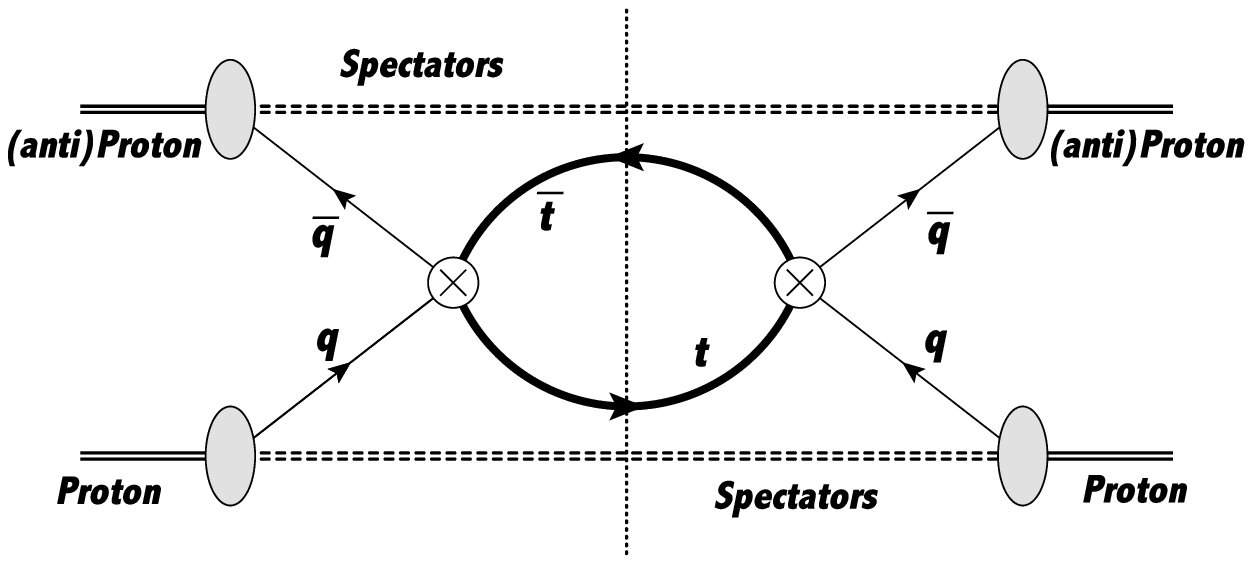} 
   \caption{Born cut diagram. The oval blobs are related to the parton distributions and the  circle
   with a cross represents the hard scattering.}
   \label{fig:Born}
\end{figure}

\begin{figure}[h]
  \centering
     \hspace{-1mm} 
   \includegraphics[width=3.5in]{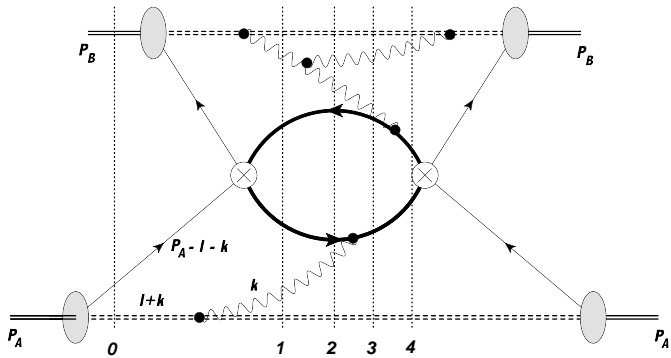} 
   \caption{
   Generic higher-order diagram. The oval blobs are related to the parton distributions and the round blob with a cross represents the hard scattering. The momenta refer to the discussion of the initial state labelled 0 in connection with Eq. (\ref{initialDA}).}
   \label{fig:generic}
\end{figure}

The basic, parton model  contribution to the cross section for pair production is represented as in Fig.~\ref{fig:Born}
in cut diagram notation, where the vertical dotted line (the ``cut") identifies the out state.
In the figure,  a $t\bar{t}$ pair with momenta $p_t$ and $p_{\bar{t}}$ emerges from a hard scattering in both the amplitude (to the left of the vertical line) and complex conjugate amplitude (to the right) in proton-(anti)proton scattering with in state momenta $p_A$ and $p_B$, and with
\bea
P^-=p_A^-+p_B^-\, .
\label{eq:P-def}
\eea
As usual, the hard scattering is initiated by ``active" partons, of flavors $a$ and $b$, whose momenta are taken proportional to the momenta of the incoming hadrons,
\bea
 \hat q^\mu &=& x_q\, p_H^\mu\ =\   x_q \left(  p_H^+,\, \frac{p_{H,\perp}^2}{2p_H^+},\,  p_{H,\perp} \right)\, ,  \quad H= A,B\, ,
\label{eq:xab-def}
\eea
where, as usual, $0< x_q\le 1$ and where generally we will use the notation $\hat q$ to refer to an on-shell momentum.
In our discussion below, we consider cut digarams like Fig.\ \ref{fig:Born}
and its generalization to higher loops, Fig.\ \ref{fig:generic} as LCOPT diagrams, in which all vertices are ordered.   The ordering in the
complex conjugate amplitude is opposite to that in the amplitude, so that the cut diagram as a whole describes 
forward scattering, with a sequence of states in the amplitude beginning with the in state of the
process and culminating with the out state, followed by a sequence of states in the complex conjugate that
take us back to the in state.

In Fig.\ \ref{fig:Born} and below we have simplified our representation by denoting the collection of spectators for $p_A$ and $p_B$ by  $l$ and $l'$, respectively, and showing them in the figure as double lines. Our arguments below will not depend on the $x^+$-ordering of the spectator interactions. Nonperturbative information, such as proton structure, is encoded in the initial
state functions, and in the distributions of spectators. In Fig.\ \ref{fig:Born} there is no rescattering of the pair with spectators,
and there is thus only a single final state, identical to the out state with the quark pair.

The generic form of higher-order cut diagrams that include soft final state interactions of the outgoing pair is illustrated by Fig.~\ref{fig:generic}. In the case shown in the figure, there are four final states, as indicated by the vertical lines.
\footnote{If a gluon momentum is collinear or hard it becomes a part of the parton distribution functions and/or a participant in
the hard scattering process.}

Combining all cuts in a partonic c.m. frame, we can write the contribution to the cross section from an arbitrary  region in momentum space, $\Pi_{ab}$ in which the hard scattering is initiated by parton $a$ from $A$ and $b$ from $B$, as  
\bea
2p_t^+\frac{d\sigma_{AB\to t\bar t+X}^{(\Pi_{ab})}}{d^3p_t} &=& 
\sum_{{\rm orderings\ T\ of}\ \Pi_{ab}}\
\int \prod_{{\rm loops}\,\{l\}}  d^2l_{\perp} dl^+\ \prod_{{\rm lines}\,\{k\}}\ \frac{\theta(k^+)}{2k^+}
\int_0^1 dx\; \delta\left( x -  \frac{x_ap_A^-+x_bp_B^-}{P^-}\right)\,
\nonumber\\ &\ & \hspace{-10mm} \times\  
{\cal I}_{ab/AB}^{(T)*}(x,q'_a,q'_b,p_A,p_B)\, {\cal F}^{(T)}_{ab}(x,x_ap_A,x_bp_B,p_t)\, {\cal I}^{(T)}_{ab/AB}(x,q_a,q_b,p_A,p_B)\, ,
\label{eq:result}
\eea
where now the sum over $x^+$ orderings and products over loops and lines 
refers to the entire cut diagram, including the final states.
We have introduced the integration variable $x$ to quantify the minus momentum available for the top pair and soft 
radiation in terms of the on-shell minus momenta of partons of momentum $q_a$ and $q_b$, whose large momentum components
are defined as in Eq.\ (\ref{eq:xab-def}) above.    Notice that the corresponding dependence
in the complex conjugate amplitude is independent, although in the limit of zero final state momentum transfer to the pair, $q_a'=q_a$ and $q_b'=q_b$.   Dependence on loop momenta $\{l\}$ is implicit. The function ${\cal I}^{(T)}_{ab/AB}$ in Eq.\ (\ref{eq:result}) 
represents the effects of {\cal all}  initial  states in the amplitude and ${\cal I}_{ab/AB}^{(T)*}$ in the complex conjugate amplitude.   
As noted above, initial states are precisely those states that do not include the top pair for the particular ordering, $T$.  The perturbative order of the ${\cal I}$s will not play a role in our arguments on final states, nor do we have to assume that we have summed over the full set of states necessary to cancel non-factoring initial state interactions \cite{Collins:1989gx,Bodwin:1984hc,Aybat:2008ct,Collinsbook}.   

The function ${\cal F}^{(T)}$ represents the product of denominators 
from the remaining, final states, which do include the quark pair, along with the
momentum-conserving delta function associated with the out state.     
We shall also include in $\cal F$ the short-distance factors that describe the production
of the top pair, which we denote by $H$ in the amplitude (to the left of the cut)
and $H^*$ to the right.   In LCOPT, these factors are given by denominators that are highly off-shell.

\subsection{Leading regions, initial state jets and final state interactions}
\label{sec:leading}

We wish to study the effects of final-state interactions 
at leading power in the large scales of the problem,
all of the order of the top mass.   
These contributions come from so-called ``leading regions" \cite{Collins:1989gx},
where in covariant perturbation theory, subsets of virtual lines are near the mass shell. 
These are regions (subspaces) where the integrands of loop momenta are singular and 
where momentum integrals are either pinched between coalescing singularities
or forced to end-points \cite{Eden-1966,Sterman:1978bj}.
In LCOPT, of course, all lines are treated as on-shell, but the characterization of regions still holds.   
In the following, we will use extensively the logarithmic nature of (gauge invariant combinations of) integrals in gauge theory
leading regions \cite{Collins:1989gx}.    This implies that a cancellation 
in an integrand at the singular surface will suppress 
the integrand near the
leading region,
making its contribution finite.   

In leading regions, a subdiagram of the full cut diagram has all loop momenta 
(including phase space loops) nearly parallel to the
incoming hadron $A$, another to hadron $B$, and another subdiagram has all line momenta nearly zero.
These are referred to respectively as jet-$A$, jet-$B$ and soft subdiagrams, which include
the ``spectator" lines of Figs.\ \ref{fig:Born} and \ref{fig:generic}.     Notice that lines of the
out state appear in the jet and soft subdiagrams in general.  
Such a  leading region contains a subspace of the total loop momentum and phase space at which all the jet and soft lines are exactly on shell. This subspace will sometimes be identified  below as its corresponding ``pinch surface" \cite{Sterman:1978bj}.
At the pinch surface, a line in the $A$ or $B$ jet takes on a momentum $\hat q$ that is exactly parallel to one of the
incoming hadrons, $H=A,B$, as defined above in Eq.\ (\ref{eq:xab-def}). Then for lines in the jet subdiagrams we expand
in terms of $+$ and $\perp$ components, 
\bea
q &=& \hat q + \delta q\, , 
\nonumber\\
\dq &\equiv& (\delta q^+, 0^-, \delta q_\perp)\, .
\eea
We emphasize again that in the frame we choose, both incoming hadrons are perpendicular to the spatial light cone direction (see Eq.\ (\ref{eq:in-momenta})). In the sense of light cone ordering, before the hard scattering the sum of all $x_a$ in the $A$ jet or $x_b$ in
the  $B$ jet is unity at the singular configuration.    After the hard scattering, the fractional momenta of
the remaining jet lines, the spectators, will add up to less than one. We will use the term spectators below to refer to final state partons with transverse momenta at the hadronic scale. Partons with  perturbative  transverse momenta but still at small angles to the incoming momenta will be referred to as part of the forward jets. The spectators are part of the forward jets, but the jets also include perturbative radiation in general.

Near the singular surface, we can expand the on-shell minus momenta of jet lines in either jet relative to their values at the pinch surface,
\bea
[q]^- &=& \frac{(x_qp_A+ \dq)_\perp^2}{2(\hat q^++\dq^+)}
\nn\\
&\equiv& [\hat q]^- +  \beta_q\cdot \dq +  \frac{1}{\hat q+}{\delta q} \cdot \lrg_q \cdot {\delta q}\ + \dots\, ,
\label{eq:qi-expand}
\eea
where we neglect terms beyond second order, and where  linear and quadratic terms are given explicitly by
\bea
\beta_q\cdot \dq  &=&
\frac{p_{A,\perp}\cdot \dq_{\perp}}{p_A^+}  - \frac{[p_A]^-}{p_A^+}\,  \dq^+
\nonumber\\
{\delta q} \cdot \lrg_q \cdot {\delta q}
&=&
\frac{ \left( \dq_{\perp}\right)^2}{2}
- \frac{\left( p_{A,\perp}\cdot \dq_{\perp}\right) \, \dq^+ }{p_A^+} 
+ \frac{[p_A]^-}{p_A^+}\, \left(\dq^+\right)^2  \, .
\eea
Equivalently, the components of the four-vector $\beta_q$  (always zero in the plus entry) are given by
\bea
\beta_q^\mu = \frac{1}{\hat q^+}\left( 0^+,\hat q^-, \hat q_\perp\right)\, .
\eea 
In fact, the combination $\beta_q\cdot \delta q$ is the on-shell value of the minus momentum for the linear eikonal propagator $1/(\hat q\cdot \delta q)$.  Thus, the expansion in $\delta q$ can be thought of as an expansion around the eikonal approximation for the heavy quarks \cite{Laenen:2010uz}. The quadratic terms in this expansion are
given by the nonzero elements of the matrix $\left({\stackrel{\leftrightarrow}{\gamma}_q}\right)_{\mu\nu}$, defined as
\bea
 \left(\lrg_q\right)_{ij} &=& \frac{1}{2}\delta_{ij}\, ,
  \quad \left(\lrg_q\right)_{++} = \frac{[\hat q]^-}{\hat q^+}\, , \quad \left(\lrg_{q}\right)_{i+} = \left(\lrg_{q}\right)_{+i} = -\, \frac{\hat q_i}{2\hat q^+}\, .
 \label{eq:nNcomponents}
 \eea

We notice that  all lines in jet $H$, $H=A,B$, for which $\hat q=x_q p_H$,  have the same $\beta_q$ and $\lrg_q$, 
 \bea
 \beta_q = \beta_{p_H} &\equiv& \beta_H\, , \quad q \in J_H\, ,
 \nn\\
 \lrg_q = \lrg_{p_H} &\equiv& \gamma_H\, , \quad q \in J_H\,, \quad H= A,B\, .
 \eea
The components $\delta q^+$ and $\delta q_{\perp}$ along with the three
components of soft loop momenta  control the contribution from each leading region.
In particular, the first-order term in the expansion of 
the momentum of a jet line
depends only on the scaleless vector $\beta_{p_H}$  and is
independent of $x_q$, while the second-order terms depend on $x_q$ only as an overall factor.

These results have immediate consequences for the initial-state light cone denominators.
Consider for example, the state $0$ in Fig.\ \ref{fig:generic}, 
which consists of only two lines, the active and a spectator parton.   The corresponding light cone denominator is 
\bea
p_A^- - [p_A-l-k]^- - [l+k]^-
= -\
\frac{1}{x_l(1-x_l)p_A^+}\, \, ({{\delta l}+k}) \cdot \lrg_{p_A} \cdot ({{\delta l}+k})\, ,
\label{initialDA}
\eea
in which the linear terms cancel. In the general case, there is a similar contribution from the $B$ jet when the state includes some of its
lines, and indeed, there is no linear dependence on spectator momenta in initial state denominators.   
 For this reason, the dependence of the initial state factors on the variable $x$ in Eq.\ (\ref{eq:result}) (the total minus momentum flowing into the hard scattering), can be absorbed into
overall factors like $x_l(1-x_l)$ in denominators like Eq.\ (\ref{initialDA}).   The resulting $x$ dependence  is hence smooth, and in fact analytic.
Now at any order, the logarithmic integrals associated with initial-state singularities involve only soft momenta and the transverse momenta of spectator lines, and are independent of the exact value of $x$.   We thus expect smooth behavior in $x$ to extend to all orders for initial states.
This will play an important role in our arguments below. In the following section, we turn to a study of final state interactions.

\section{Cancellation of final state interactions
in single-particle inclusive cross sections}
\label{sec:NLO}

In this section we discuss the sum over the choice of out state among the final states of the cut diagram 
represented in Eq.\ (\ref{eq:result}).
Taken together these factors  may be represented
in the general case of $S$ final states as 
\begin{eqnarray}
{\cal F}^{(T)} &=& \sum_{j=1}^S\ \int d^3p^{(j)}_t\ \left(\prod _{i'=j+1}^S 
\frac{1}{P^- - s_{i'}-i\epsilon}\right) 2\pi\delta\left(P^- - s_{j}\right) \left(\prod _{i=1}^{j-1} \frac{1}{P^- - s_{i}+i\epsilon}\right)\ \delta^3( p_t-p_t^{(j)}) \, ,
\nonumber\\
&\ & \hspace{10mm} \times\ H_{ab}^*(x_ap_A,x_bp_B,p_t)\, H_{ab}(x_ap_A,x_bp_B,p_t)\, ,
\label{eq:F_fixed_cut}
\end{eqnarray}
where, again, $s_j$ is the sum of on-shell minus momenta in the out state,
and similarly for the final states in the amplitude and complex conjugate.
The hard scattering functions $H$ and $H^*$ depend only weakly on soft momenta,
and we can consider them as functions only of the on-shell active parton
and the observed top momenta.

In our analysis of the nearly on-shell light cone denominators, of course, we
must keep track of the top
quark momentum in each state.    
The factor $\delta^3( p_t-p_t^{(j)})\equiv \delta(p_t^+-p_t^{(j)}{}^+)\delta^2(p_{t,\perp}-p^{(j)}_{t,\perp})$
 fixes the top quark momentum that appears in out state $j$ to equal the prescribed value $\vec p_t$.
 For each choice of out state, $j$ we treat the momentum of the top in that state, $p^{(j)}_t$  as a loop 
 momentum  that passes through the
top and anti-top lines and the hard scatterings only.   The momenta of top lines in all other final states is then fixed by $p_t$
 and the sum of soft gluon emission and absorption. In the sum over $j$, we let each final state play the role of the out state in turn.
Again, we suppress numerator factors, and because the choice of light cone order, $T$ will be fixed for our argument, we will suppress it as well below.

\subsection{Lowest order}
\label{subsec:loworder}

To illustrate the mechanism of cancellation, 
we begin with final-state hard parton (top)-spectator interactions in single particle inclusive observables at first order. 
As we will show in the following subsection, the generalization of the proof of cancellation
to all perturbative orders is straightforward. 

\begin{figure}[t]
  \centering
     \hspace{-1mm} 
   \includegraphics[width=0.46\textwidth]{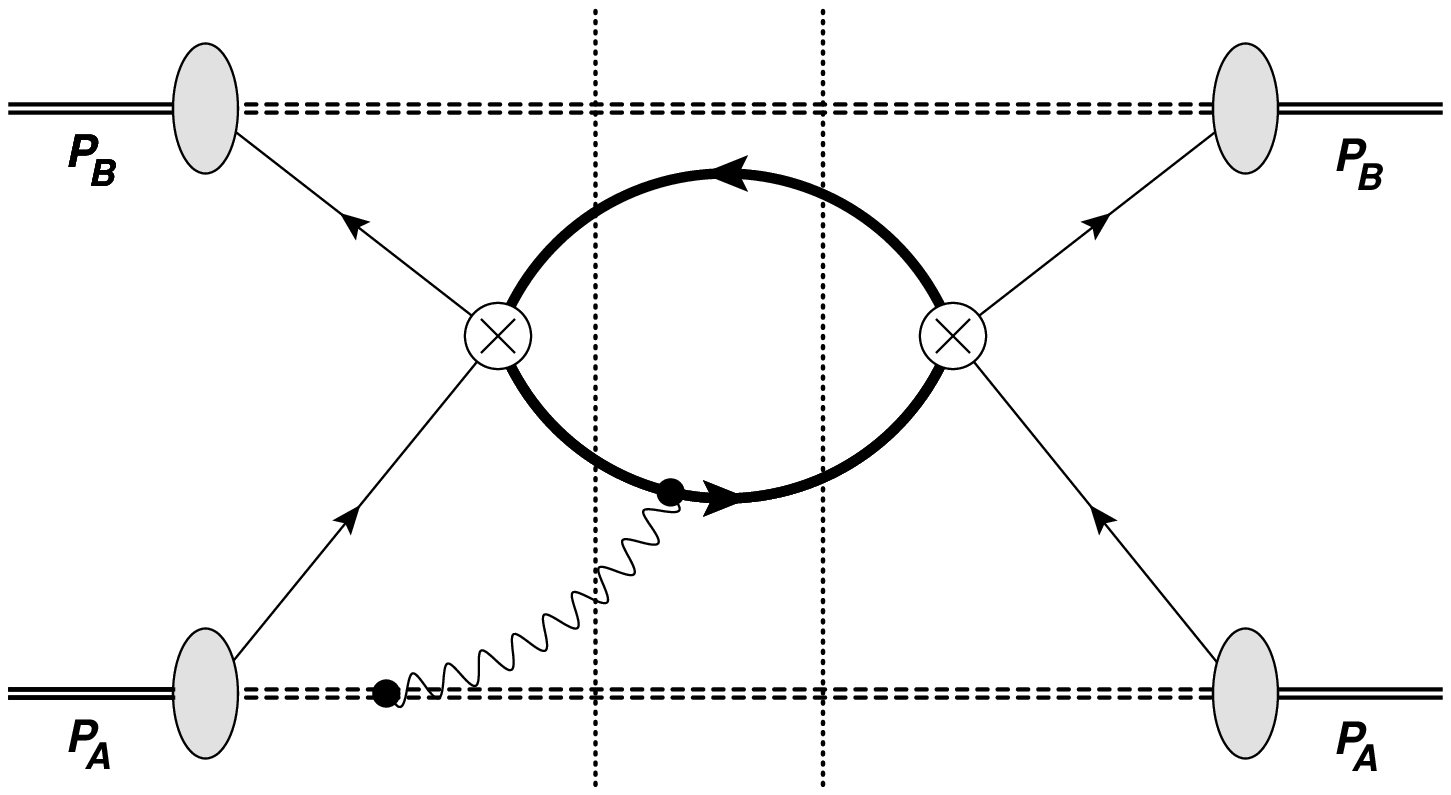}
   \hfill
   \includegraphics[width=0.46\textwidth]{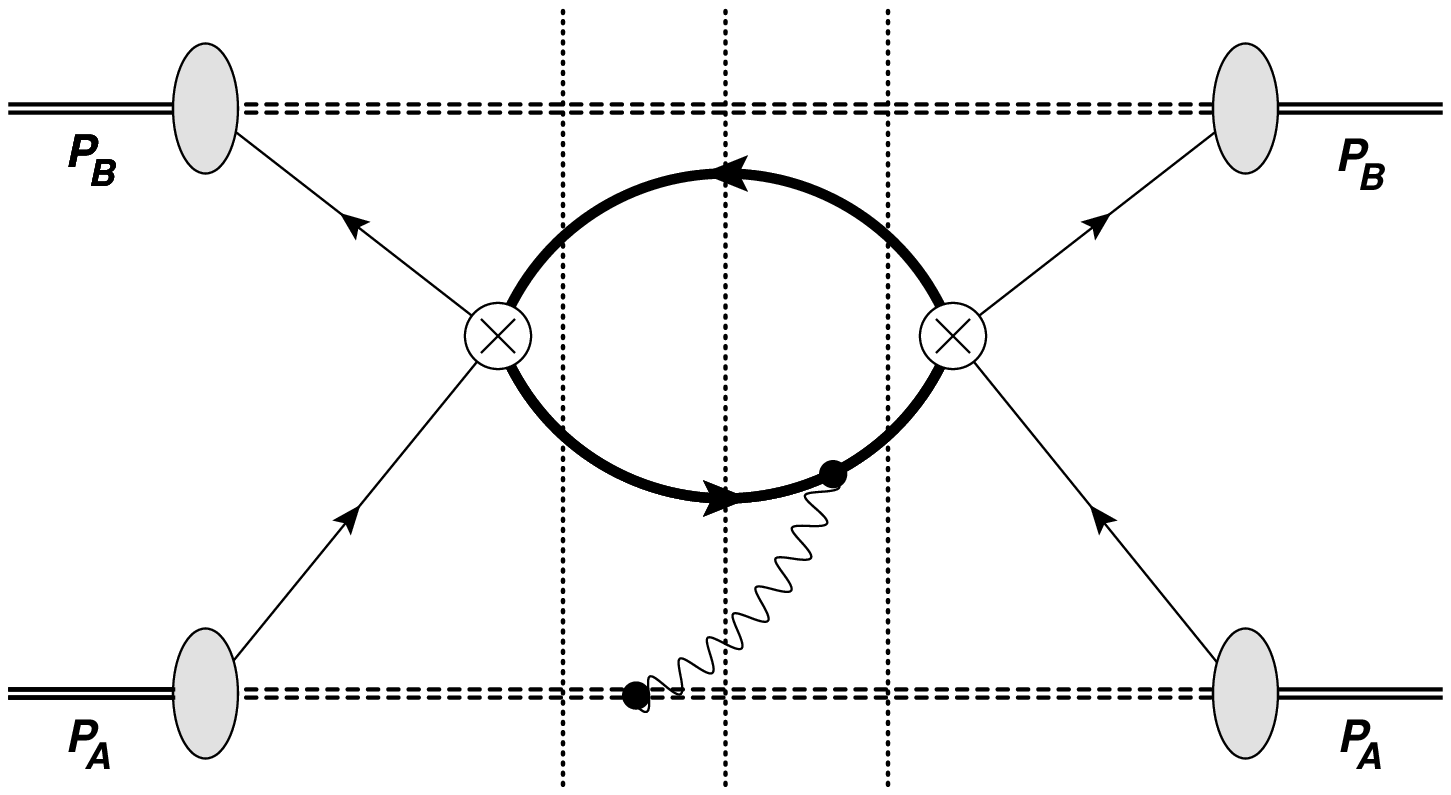} \\
   \hbox{\hskip 1.5 in (a) \hskip 3.65 in (b) }
   \caption{The LCOPT diagrams with first-order final state interactions discussed in the text.}
   \label{fig:NLO}
\end{figure}

First order soft final-state hard parton-spectator interactions are shown in Fig.~\ref{fig:NLO}.   In these ordered diagrams, a single soft gluon is emitted from the spectators of line $p_A$ and absorbed by the top quark.    In addition to these two diagrams there are also their complex conjugates, exchanges between the $t$ quark and the $p_B$-jet spectators, diagrams where $t$ is exchanged with $\bar{t}$, and also diagrams where the gluon is emitted from an ``active" line.   The reasoning in these cases is equivalent.  

In  Fig.~\ref{fig:NLO}a, the soft gluon is emitted from an initial state; in Fig.\ \ref{fig:NLO}b from a final state.    The former case has two final states, the latter three.
We are interested in leading regions that involve soft gluon exchange, and we should note that the range  in gluon momenta with leading-power behavior depends on the nature of the final state process.  For example, in diagrams like those in Fig.\ \ref{fig:NLO} the wide-angle radiation of an on-shell gluon from a spectator with longitudinal momentum $l^-\sim xP$ and  transverse momentum  $\langle l_T \rangle$ (relative to the beam axis) is  leading power only for very soft momenta, $k^\pm\sim k_T \le  \langle l_T \rangle^2/ xP$ \cite{Qiu:1990xy,Collins:1982wa}.  In contrast, an off-shell gluon exchange that mediates the elastic scattering of the top quark by a spectator is leading power all the way to the scale of the spectator's transverse momentum, $k_T \sim \langle l_T \rangle$.   Naturally such scattering processes have greater potential for phenomenological relevance.  The arguments we give below cover both of these cases, however, and we will generally assume that soft gluon exchange involves momentum transfers up to the scale of the transverse momenta of spectators.

\begin{figure}[t]
  \centering
     \hspace{-1mm} 
   \includegraphics[width=0.46\textwidth]{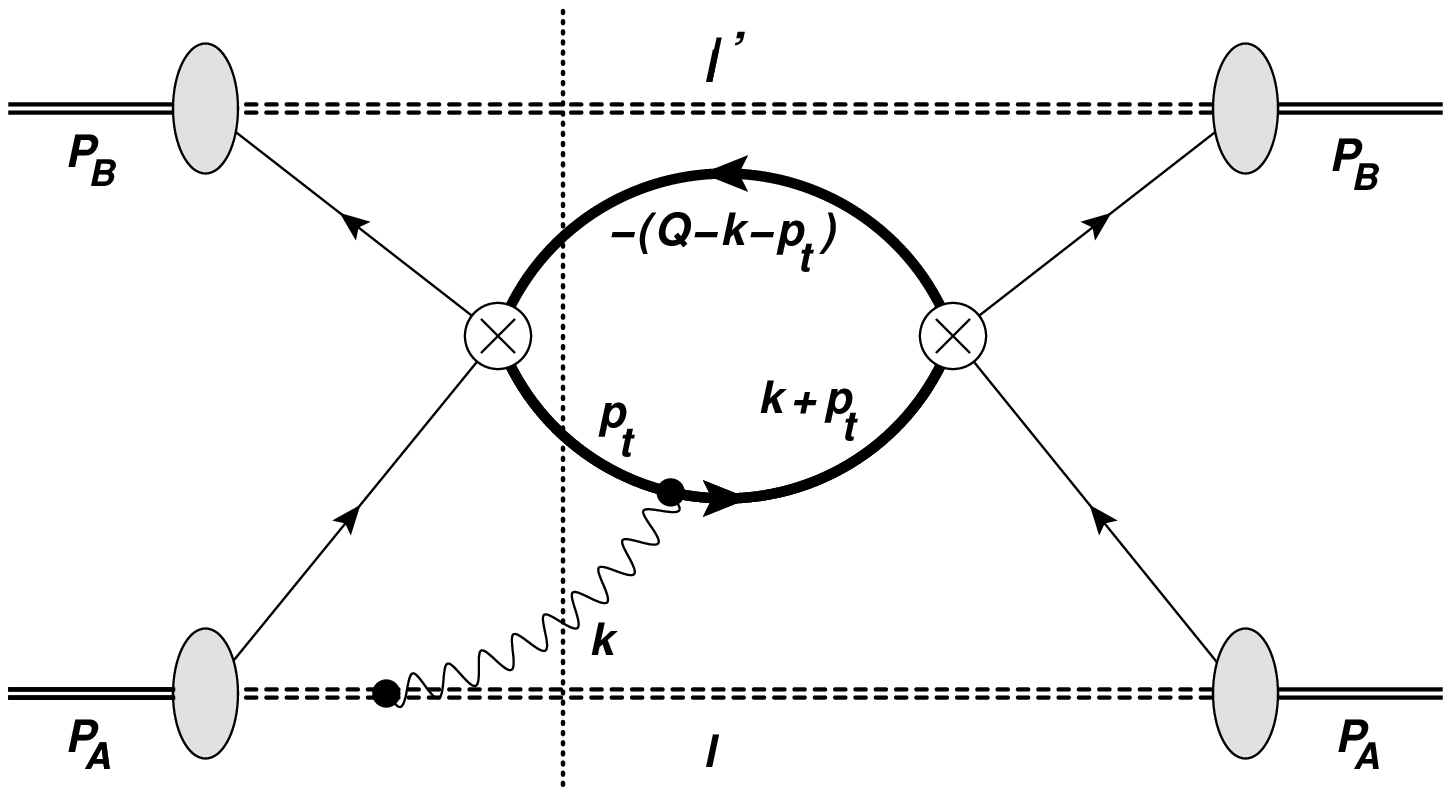}
   \hfill
   \includegraphics[width=0.46\textwidth]{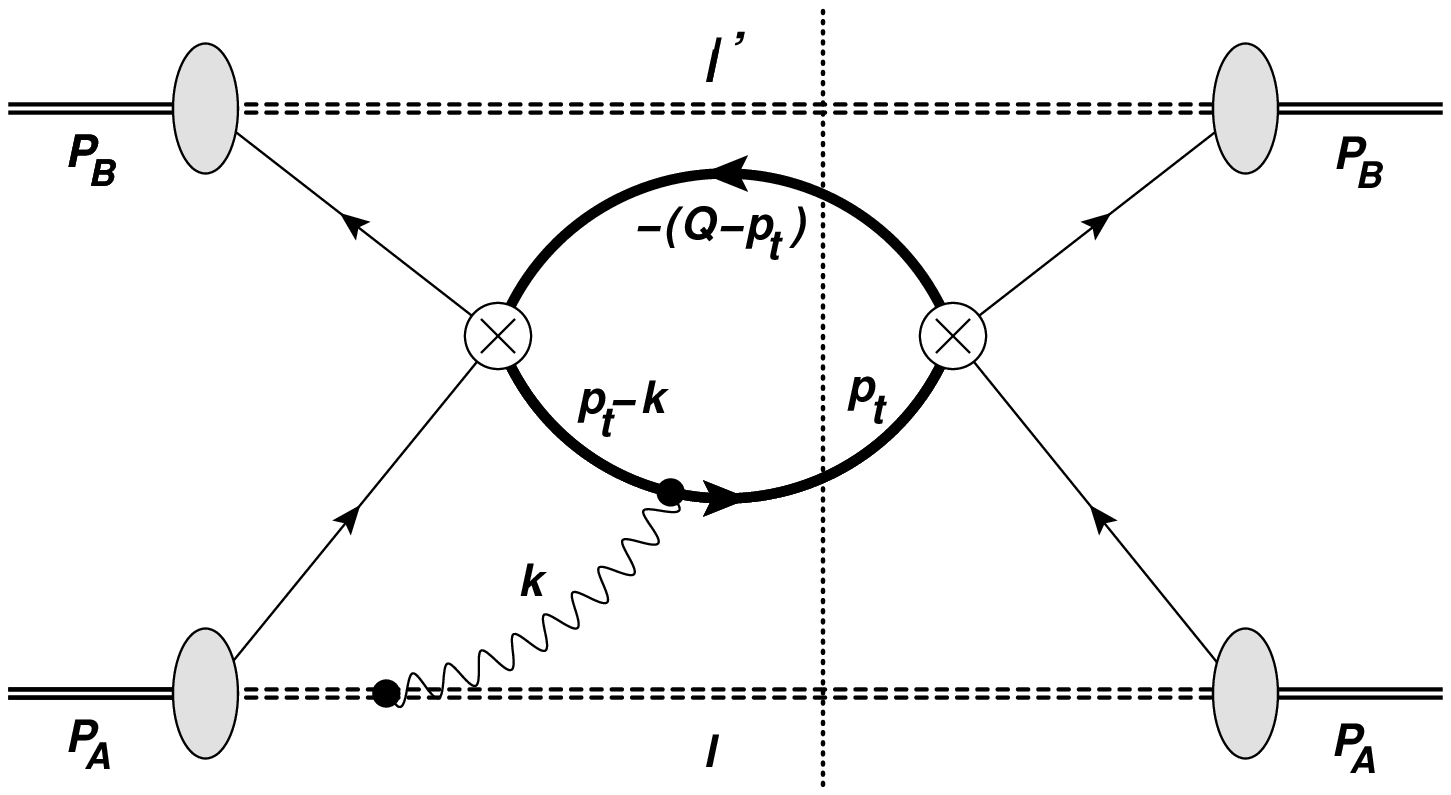} 
   \caption{The two cuts of the left diagram on Fig.~\ref{fig:NLO}. The relevant three-momenta are shown and $Q \equiv P_A+P_B-l-l'$. }
   \label{fig:NLO-cuts}
\end{figure}

We begin with the simplest case of Fig.~\ref{fig:NLO}a, which has only two final states.
The sum over the two choices of out state (``cuts") in Fig.~\ref{fig:NLO}a is shown explicitly in Fig.\ \ref{fig:NLO-cuts},
with assigned momenta for the pair. After using the three-dimensional momentum delta function to do the $p_t^{(j)}$ integral,
the quantity ${\cal F}$ defined in Eq.~(\ref{eq:F_fixed_cut}) can be written, here with $S=2$ as
\begin{eqnarray}
{\cal F} &=& \left [  
2\pi\delta\left(D_2^{(2)}\right) {1\over D_1^{(2)}+i\epsilon} 
+ 
{1\over D_2^{(1)}-i\epsilon} 2\pi\delta\left(D^{(1)}_1\right)  \right]  H_{ab}^*(x_ap_A,x_bp_B,p_t)\, H_{ab}(x_ap_A,x_bp_B,p_t) \, ,
\label{eq:Soriginal}
\end{eqnarray}
where we denote by $D_i^{(j)}$ the minus momentum deficit (as in Eq.\ (\ref{eq:F_fixed_cut})) of final state $i$ when $j$ is the out state. When $i=j$, $D_j^{(j)}$ is the argument of the delta function in the corresponding term. 
For convenience, we define the total momentum flowing into the hard scattering in the amplitude by
 \bea
 Q \equiv p_A+p_B-l-l'\, ,
 \label{eq:Q-def2}
 \eea
where $l$ and $l'$ represent all spectators at the pinch surface in the first (final) state in the amplitude after the hard scattering.   We
do not include in $Q$ the momenta of  soft lines like $k$ in Fig.\ \ref{fig:NLO-cuts}, which carry momenta between spectators and the top pair in the final state. In these terms, the functions $D_i^{(j)}$ for Eq.\ (\ref{eq:Soriginal}) can be written as
\begin{eqnarray}
D_1^{(1)} &=& x P^- - [Q-p_t-k]^- - [p_t]^- + d_1\, , \nonumber\\
&=& x P^- - [Q-p_t]^- - [p_t]^- +   \beta_{Q-p_t}\cdot k - \frac{1}{(Q-p_t)^+}k\cdot \lrg_{Q-p_t}\cdot k + d_1+ \dots \, , \nonumber\\
D_1^{(2)} &=& x P^- - [Q-p_t]^- - [p_t-k]^- + d_1 \, , \nonumber\\
&=& x P^- - [Q-p_t]^- - [p_t]^- +   \beta_{p_t}\cdot k - \frac{1}{p_t^+} k\cdot \lrg_{p_t} \cdot k+ d_1+ \dots \, , \nonumber\\
D_2^{(1)} &=& x P^- - [Q-p_t-k]^- - [p_t+k]^- + d_2\, , \nonumber\\
&=& x P^- - [Q-p_t]^- - [p_t]^- +  \left( \beta_{Q-p_t}  -\beta_{p_t} \right) \cdot k +\frac{1}{(Q-p_t)^+}k\cdot \lrg_{Q-p_t}\cdot k  - \frac{1}{p_t^+} k\cdot \lrg_{p_t} \cdot k 
+d_2+ \dots \, , \nonumber\\
D_2^{(2)} &=& x P^- - [Q-p_t]^- - [p_t]^- + d_2 \, .
\label{eq:AandD}
\end{eqnarray}
In the second equalities for each of the first three $D_i^{(j)}$ we have expanded to second order in soft momentum $k$,
following Eq.\ (\ref{eq:qi-expand}). For each denominator, we have added and subtracted the term $xP^-$, defined
in Eq.\ (\ref{eq:result}), absorbing the term $-xP^-$,
into $d_1$ and $d_2$, which depend only on the
final state $i$ and not on the choice of out state $j$.   
The $d_i$ in Eq.\ (\ref{eq:AandD}) also depend on 
the details of the spectator and soft lines, whether or
not connected directly to the top loop.    
Our arguments will not depend on their explicit form.   To give an example, however, we can treat
the double-dashed lines of Fig.\ \ref{fig:NLO} as single
spectators, of momenta $l$ and $l'$, which gives
\bea
d_1 &=& (1-x)P^-   -[k]^- -[l]^--[l']^- \, ,
\nonumber\\
d_2 &=& (1-x)P^- -[l]^--[l']^-\, .
\label{eq:deltaD-0}
\eea
We expand jet line momenta $l$ and $l'$ about the pinch surface, where $Q=x_aP_A+x_bP_B$ in Eq.\ (\ref{eq:Q-def2}).   The expansion then follows Eq.\ (\ref{eq:qi-expand}), with 
\bea
[\hat l]^- &=& (1-x_a)p_A^-\, ,
\label{eq:hat-llprime}\\
\nn
[\hat l']^- &=&(1-x_b)p_B^-\, ,
\eea
where again $x_h$ is the fractional momentum of the
active parton from hadron $H=A,B$ at the pinch surface.    
To second order, as
given in Eq.\ (\ref{eq:qi-expand}), we find
\bea
d_1 &=& -\  \frac{k_\perp^2}{2k^+} 
- \beta_{p_A}\cdot \delta l + \frac{\delta l\cdot \lrg_{p_A}\cdot \delta l}{l^+} - \beta_{p_B}\cdot \delta l'  + \frac{\delta l'\cdot \lrg_{p_A}\cdot \delta l'}{l'{}^+}\, ,
\nonumber\\
d_2 &=& 
-\ \beta_{p_A}\cdot \delta l + \frac{\delta l\cdot \lrg_{p_A}\cdot \delta l}{l^+} - \beta_{p_B}\cdot \delta l' + \frac{\delta l'\cdot \lrg_{p_A}\cdot \delta l'}{l'{}^+}\, .
\label{eq:deltaD}
\eea
From Eqs.\ (\ref{eq:Q-def2}) and (\ref{eq:hat-llprime}) we see that within each of the terms of (\ref{eq:Soriginal}), the large (that is, order $m_t$) terms cancel, and the uncut final state denominator is independent of $x$, and is of order of the components of vector $k^\mu$.  Thus, the contributions from ${\cal F}$ to the corresponding cut diagrams are order $1/k$, where $k$ will stand collectively
for the terms $[k]^-$, $\beta_A\cdot l$, $\beta_B\cdot l'$. We will  show that this $1/k$ behavior cancels after the sum over the
two cuts. 

To exhibit the cancellation of the singular, $1/k$ behavior just identified in final state interactions, we will apply the relation
\begin{equation}
2\pi\delta(y) =  i\, \left( \frac{1}{y+i\epsilon} - \frac{1}{y-i\epsilon} \right)
\label{eq:delta-function}
\end{equation}
to Eq.~(\ref{eq:Soriginal}). This results in four terms, each a product of two propagators,
\begin{eqnarray}
-i {\cal F}
&=& \left [ {1\over D_2^{(2)}+i\epsilon}\times {1\over D_1^{(2)}+i\epsilon} 
\quad -\ {1\over D_2^{(2)}-i\epsilon} \times {1\over D_1^{(2)}+i\epsilon}  +  {1\over D_2^{(1)} - i\epsilon}\times {1\over D_1^{(1)} + i\epsilon} 
    - {1\over D_2^{(1)}-i\epsilon}\times {1\over D_1^{(1)} - i\epsilon}
    \right ]
\nonumber\\
&\ & \hspace{0mm} \times\, H_{ab}^*(x_ap_A,x_bp_B,p_t)\, H_{ab}(x_ap_A,x_bp_B,p_t) \, .
\label{eq:4terms}
\end{eqnarray}
Of these four terms, the first and fourth have both $i\epsilon$ prescriptions the same
in their denominators.    These products of final-state
denominators therefore do not produce a pinch in the variable $x$, and since $x$ dependence is otherwise
analytic in the leading region we can deform the $x$ contour away from points where the denominators
would otherwise vanish.     When $x$ changes by any finite amount,
it forces these denominators off-shell by an amount of order $P^-$, and their contributions
can be absorbed in the hard scattering function.

In the remaining two terms of Eq.\ (\ref{eq:4terms}) there are denominators with opposite $i\epsilon$'s,
so that in these terms the $x$ integral is pinched in general.    What we will now show is that singular behavior associated with these
terms cancels.    The mechanism of cancellation will be readily generalized to arbitrary order.

Neglecting the nonsingular terms we have
\begin{eqnarray}
- i{\cal F}
&=& \left[ {1\over D_2^{(1)} - i\epsilon}\times {1\over D_1^{(1)} + i\epsilon} -\ {1\over D_2^{(2)}-i\epsilon} \times {1\over D_1^{(2)}+i\epsilon} \right] H_{ab}^*(x_ap_A,x_bp_B,p_t)\, H_{ab}(x_ap_A,x_bp_B,p_t)
\nonumber\\
&=&  \left[ {1\over D_2^{(2)} + \left(D_2^{(1)}-D_2^{(2)}\right)-i\epsilon} \times  {1\over D_1^{(2)} + \left(D_1^{(1)}-D_1^{(2)}\right)+i\epsilon}   
-  {1\over D_2^{(2)}-i\epsilon} \times {1\over D_1^{(2)}+i\epsilon} \right]
\nonumber\\
&\ & \hspace{5mm} \times\
H_{ab}^*(x_ap_A,x_bp_B,p_t)\, H_{ab}(x_ap_A,x_bp_B,p_t) \, ,
\label{eq:Fprime-def}
\end{eqnarray}
where the trivial rewriting of the second form shows manifestly that
{\it if} $D_i^{(1)}$ equaled $D_i^{(2)}$ for $i=1,2$, the two terms would cancel identically.
The differences, which are linear in the soft momentum $k$,  can be read off from Eq.\ (\ref{eq:AandD}).
In the notation of Eq.\  (\ref{eq:qi-expand}), we find
\bea
D_1^{(1)}-D_1^{(2)} &=& - [p_t]^- - [Q-p_t-k]^- + [p_t-k]^- + [Q-p_t]^- 
\nonumber\\
&=&   (\beta_{Q-p_t} - \beta_{p_t})\cdot k + \frac{k \cdot (\lrg_{p_t})\cdot k}{2p_t^+} - \frac{k \cdot (\lrg_{Q-p_t}) \cdot k}{(Q-p_t)^+} + \dots
\nonumber\\
D_2^{(1)}-D_2^{(2)} &=& - [p_t+k]^- - [Q-p_t-k]^- + [p_t]^- + [Q-p_t]^-
\nonumber\\
&=& (\beta_{Q-p_t} - \beta_{p_t})\cdot k -  \frac{k \cdot(\lrg_{p_t})\cdot k}{2p_t^+} - \frac{k\cdot (\lrg_{Q-p_t})\cdot k}{(Q-p_t)^+} + \dots\, ,
\label{eq:first-differences}
\eea
both independent of $x$. To compensate for these differences we will again appeal to our observation above that the remainder of the diagram
has a smooth dependence on the collective parton fraction $x$. Thus, up to corrections suppressed by $1/P^-$
we may perform a small shift, $x \rightarrow x + \delta x$ in the first term in the second equality of Eq.\ (\ref{eq:Fprime-def}), where 
\bea
\delta x =\frac{ ( \beta_{Q-p_t} - \beta_{p_t}) \cdot k }{P^-}\, ,
\label{Deltax}
\eea
which is power suppressed in the hard scale.    We then find
\begin{eqnarray}
-i{\cal F}
 = \left[{1\over D_2^{(2)} + \hat\Delta^{(2)}_2 -i\epsilon }\ {1\over D_1^{(2)} + \hat\Delta^{(2)}_1+i\epsilon} - 
 {1\over D_2^{(2)}-i\epsilon} \ {1\over D_1^{(2)} +i\epsilon} \right] H_{ij}^*(x_ap_A,x_bp_B,p_t)\, H_{ij}(x_ap_A,x_bp_B,p_t)\, ,
\label{eq:S}
\end{eqnarray}
where $\hat\Delta^{(2)}_1$ and $ \hat\Delta^{(2)}_2$ are both of order $k^2$. They can be read off from Eq.\ (\ref{eq:first-differences}), and
\bea
 \hat\Delta^{(2)}_1\ -\ \hat\Delta^{(2)}_2  
\ =\ 2\ \frac{k \cdot (\lrg_{p_t}) \cdot k}{p_t^+} + \dots\, .
\label{eq:Delta-2ndorder}
\eea
We conclude that because pinches of the $x$ integral are found only in ${\cal F}$, the entire term integral 
is suppressed  for fixed values of the soft gluon momentum $k$.
 We recall again that (gauge invariant) perturbative contributions to single-particle inclusive cross section are at worst logarithmically divergent \cite{Nayak:2005rt}.  Thus, cancellation of the leading power in $k$ will lead to a finite integral.   Under these circumstances, the integral over momentum $k$ is dominated by $k\sim m_t$, and to fixed order in perturbation theory the gluon $k$, and every line connected to it can be absorbed into the hard scattering.   
 The integration region $k\rightarrow 0$ appears only as the tail of a finite integral, and its contribution vanishes for $m_t\rightarrow 0$.

In summary, to lowest order in final state interactions we have seen that cancellation is manifest once we neglect terms that are quadratic in soft momenta compared to those that are linear in each of the light-cone denominators. Alternatively, if we expand the expression (\ref{eq:S}) for ${\cal F}$ in powers of $ \hat\Delta^{(2)}_n/D_n^{(2)} \sim k,~ n=1,2$, the leading terms cancel and the integral is  finite at the pinch surface.

\subsection{Final state interactions at arbitrary order}
\label{sec:allorders}

We now generalize to arbitrary order the mechanism of cancellation found above at fixed order and with the minimum number of final states. For a generic diagram with $S$ states, as the one illustrated by Fig.~\ref{fig:generic} and 
Eq.\ (\ref{eq:Soriginal}), the function ${\cal F}$ can be written 
as a sum over out states $j$,
\begin{eqnarray}
{\cal F} &=& \sum_{j=1}^S \left(\prod _{i'=j+1}^S \frac{1}{D_{i'}^{(j)}-i\epsilon}\right) 2\pi\delta\left(D_{j}^{(j)}\right) \left(\prod _{i=1}^{j-1} \frac{1}{D_{i}^{(j)}+i\epsilon}\right)\ 
 \nn\\
 &\ & \hspace{20mm}
\times H_{ab}^*(x_ap_A,x_bp_B,p_t)\, H_{ab}(x_ap_A,x_bp_B,p_t)
\nonumber\\
&=&  i\left[ \sum_{j=1}^S \left(\prod _{i'=j+1}^S \frac{1}{D_{i'}^{(j)}-i\epsilon}\right)  \left(\prod _{i=1}^{j} \frac{1}{D_{i}^{(j)}+i\epsilon}\right) 
- \sum_{j=1}^S \left(\prod _{i'=j}^S \frac{1}{D_{i'}^{(j)}-i\epsilon}\right)  \left(\prod _{i=1}^{j-1} \frac{1}{D_{i}^{(j)}+i\epsilon}\right) \right]
\nonumber\\
&\ & \hspace{20mm} \times\ H_{ab}^*(x_ap_A,x_bp_B,p_t)\, H_{ab}(x_ap_A,x_bp_B,p_t)\, .
\label{eq:F-all-order}
\end{eqnarray}
In the second equality we have used the delta function identity Eq.\ (\ref{eq:delta-function}).  
Again we work with a fixed ordering, and hence suppress the ordering label, $T$.

As above, we will expand the denominators $D_i^{(j)}$ around an arbitrary pinch surface.   
Near the pinch surface, each final state $i$ will contain a top quark, an antitop quark,
lines parallel to $p_A$ (the $A$-jet), lines parallel to $p_B$ (the $B$-jet) and soft lines.   
There may also be light parton final state jets, but for the moment we neglect this possibility.

In the single-particle inclusive cross section, the top quark is fixed to  momentum $p_t$ 
when state $i=j$, the out state in Eq.(\ref{eq:F-all-order}).   The top quark momentum in any
other final state $i$ then depends on the choice of out-state $j$, 
because the flow of soft momenta must be adjusted as the choice of out state $j$ changes.  
As above, we choose to sum over states at fixed on-shell momenta for all soft and jet lines,
adjusting the flow of soft momenta only within the top quark loop.   
We denote the resulting top quark momentum for final state $i$ with out state $j$ by $p^{(j)}_{t,i}$,
where $p^{(j)}_{t,j}=p_t$.    Similarly, antitop momenta will be denoted by $p^{(j)}_{\bar t,i}$.

In addition to the top pair for each final state, $i$, the momenta of soft lines are denoted collectively by $k,\ k\in i$, and
the momenta of
lines in the $A$ and $B$ jets denoted collectively by $l=x_lp_A+\delta l,\ l\in i$ and $l'=x_{l'}p_B+\delta l',\ l'\in i$.

In the notation just described, a generic final state denominator can be written as
\begin{eqnarray}
\quad D_i^{(j)} &\equiv& P^- - [p^{(j)}_{t,i}]^- - [p^{(j)}_{\bar t,i}]^ -   - d_i(\{k\},\{l\},\{l'\}) \,,
\label{eq:Dij-def}
\end{eqnarray}
where as above $P^-=p_A^-+p_B^-$ and where
the function $d_i(\{k\},\{l\},\{l'\})$ contains the on-shell minus momenta of the soft lines and collinear
spectator lines in final state $i$,
\begin{eqnarray}
d_i(\{k\},\{l\},\{l'\}) &=& \sum_{k\in i} [k]^- + \sum_{l\in i} [x_lp_A+\delta l]^- + \sum_{l'\in i} [x_{l'}p_B+\delta l']^-\, ,
\end{eqnarray}
which we can expand about the pinch surface, as in Eqs.\ (\ref{eq:deltaD-0}) and (\ref{eq:hat-llprime}),
\bea
d_i(\{k\},\{l\},\{l'\}) &=&
d_i(\{k=0\},\{l=x_lP^-\},\{l'=x_{l'}P^-\})
+  \hat d_i(\{k\},\{\delta l\},\{\delta l'\})
\nonumber\\
&\equiv& 
(1-x)P^- + \;  \hat d_i(\{k\},\{\delta l\},\{\delta l'\}) \, .
\eea
Using this expansion near the singular point in (\ref{eq:Dij-def}),  we then have a direct generalization of Eq.\ (\ref{eq:AandD}),
\begin{eqnarray}
D_i^{(j)}  &=& xP^- - [p^{(j)}_{t,i}]^-  - [p^{(j)}_{\bar t,i}]^ - - \hat d_i(\{k\},\{\delta l\},\{\delta l'\})\, .
\label{eq:Dijexpand}
\end{eqnarray}
As in Eq.\ (\ref{eq:AandD}), the {\it only} dependence of $D^{(j)}_i$, Eq.\ (\ref{eq:Dijexpand})
on the choice of out state is in the top quark and antiquark momenta $p^{(j)}_{t,i}$ and $p^{(j)}_{\bar t,i}$.   
Also as in the previous subsection, we denote by $k$ the collection of all soft gluon momenta $k$,
and collinear $\delta l$ and $\delta l'$.  In this notation, all the  $D^{(j)}_i$ are linear in $k$ for $k\rightarrow 0$
as above.

For $i$ the out state, that is $i=j$ in our notation, $p_{t,j}^{(j)}=p_t$ is the momentum that defines the inclusive cross section.
This is the direct generalization of the lowest-order procedure above, and will lead directly
to manifest cancellation of final state interactions, by considering the differences between denominators,
\bea
D_i^{(j+1)}-D_i^{(j)} &=& - [p^{(j+1)}_{t,i}]^-  - [p^{(j+1)}_{\bar t,i}]^ - + [p^{(j)}_{t,i}]^-  + [p^{(j)}_{\bar t,i}]^ -\, .
\label{eq:D-differences}
\eea
We consider, then, the dependence of the top and antitop momenta in an arbitrary final state $i$ on the choice of out state, $j$.

Consider now final state, $i\ne j$, which may be in the amplitude ($i<j$) or the
complex conjugate $(i>j)$.   Given the momentum routing we have chosen, a top quark line momentum in either the amplitude 
or complex conjugate will differ from $p_t$ by those soft momenta that are either emitted or absorbed by the top between state $i$ and state $j$.
We can then write
the momentum of the top quark in state $i$ that appears in Eq.\ (\ref{eq:D-differences}) as
\begin{eqnarray}
[p_{t, i}^{(j)}]^- &=& [\, p_t - \sum_{l=i}^{j-1} \sigma_l k_l\, ]^-  ~~ {\rm for}~\ i\le j\, , \quad [\, p_{t, i}^{(j)}\, ]^- = [\, p_t + \sum_{l= j}^{i-1} \sigma_l k_l\, ]^- ~~{\rm for} ~\ i>j\, ,
\label{eq:t-flow}
\end{eqnarray}
where $\sigma_l=1$ if $k_l$ is absorbed by the top quark between state $l$ and state $l+1$, and $\sigma_l=-1$ if $k_l$ is emitted from the top quark between states $l$ and $l+1$.   When the states $l$ and $l+1$ are separated by an interaction that does not involve the top quark, $\sigma_l=0$.    We note that in LCOPT the term ``emitted" refers to a gluon whose momentum $k_l$ flows forward from the vertex that separates states $l$ and $l+1$ into state $l+1$, while ``absorbed" implies that the gluon flows forward into the vertex from state $l$.

The situation for antitop momenta is similar, but in this case the momentum in state $i$ equals a value that
is independent of the choice of out state plus a correction due to the rerouting of soft momenta as we change $j$.
All soft momenta that are emitted or absorbed from the top line after out state $j$ are routed through the antitop line,
while those that are emitted before $j$ are not.       
The $j$-independent part of $p_{\bar t, i}^{(j)}$
includes soft line momenta that attach directly
to the antitop line, and dependence on all other soft and collinear loops that do not attach to the
top quark, and which are held fixed as we change the choice of out state $j$.
In summary, we denote the $j$-independent part of the antitop momenta by $\tilde p_{\bar t,i}$, and in these terms we have
\begin{eqnarray}
[p^{(j)}_{\bar t,i}]^- &=& [\, \tilde p_{\bar t,i}  - \sum_{l\ge j} \sigma_l k_l\, ]^-   ~~ {\rm for\ all}~\ i\, , 
\label{eq:tbar-recoil}
\end{eqnarray}
where again $\sigma_l=+1$ if $k_l$ is absorbed by the top quark and $\sigma_l=-1$ if $k_l$ is emitted from the top quark.
We can now evaluate the change in light cone denominators for a change in 
out
state, Eq.\ (\ref{eq:D-differences}).

We observe that $D^{(j+1)}_i \ne D_i^{(j)}$ only when $\sigma_j\ne 0$, that is, when  the vertex that separates state $j+1$ from state $j$ 
is the final interaction of the top quark before the out state.
Otherwise the top and antitop momenta remain unchanged, and
$D^{(j+1)}_i - D_i^{(j)}=0$ identically for all $i$.   In the case when the soft gluon-top quark interaction
passes from the amplitude to the complex conjugate, we expand Eq.\ (\ref{eq:D-differences}) using (\ref{eq:qi-expand}) to find
\begin{equation}
 D_i^{(j+1)}-D_i^{(j)}\  =\ \sigma_j\, \left(\beta_{p_t} - \beta_{Q-p_t}\right) \cdot k_j  \ +\ \hat\Delta_i^{(j)} \, ,
\label{eq:Dij-diff-2}
\end{equation}
where the terms linear in the soft momentum are independent of $i$.
The remaining contributions, $\hat\Delta_i^{(j)} \sim  {\cal O}(k^2/m_t)$
are quadratic in all soft momenta connected to the top loop, generalizing the lowest order result in Eq.\ (\ref{eq:Delta-2ndorder}). 
The linear terms Eq.~(\ref{eq:Dij-diff-2}) are independent of the choice of final state $i$ for any fixed out state $j$. 
 
We now return to the analysis of the general final state factor in Eq.\ (\ref{eq:F-all-order}), and
 notice that as in Sec.\ \ref{sec:NLO}, two terms, $j=1$ in the first sum of the expanded form and $j=S$ in the second, have all
 denominator poles
 on the same side of the $x$ contour.    These terms are suppressed by a power of the 
 overall energy scale, $P^-$ by the same contour deformation argument for parameter $x$ as above.
Again following the argument of Sec.\ \ref{sec:NLO} we combine the remaining $2S-2$ terms into a sum of $S-1$ pairs
with equal numbers of denominators with $+i\epsilon$ and equal numbers of denominators with $-i\epsilon$.
 To implement this step, we simply modify the summation variable $j\to j+1$ in the second sum of the second equality
 of Eq.\ (\ref{eq:F-all-order})
  and then combine terms in the form of Eq.~(\ref{eq:Fprime-def}) to derive
\begin{eqnarray}
-i{\cal F} &=& \sum_{j=1}^{S-1} {\cal F}^{(j)} + \dots  \, ,
\label{eq:Fj-def}
\end{eqnarray}
where we suppress terms in which the $x$ integral can be deformed and where
\begin{eqnarray}
-i{\cal F}^{(j)} &=& \left[\, \left(\prod _{i'=j+1}^S \frac{1}{D_{i'}^{(j)}-i\epsilon}\right) \left(\prod _{i=1}^j \frac{1}{D_{i}^{(j)}+i\epsilon}\right)  
-  \left(\prod _{i'=j+1}^S \frac{1}{D_{i'}^{(j+1)}-i\epsilon}\right) \left(\prod _{i=1}^j \frac{1}{D_{i}^{(j+1)}+i\epsilon}\right) \right]
\nn\\
&\ & \hspace{5mm} \times\
H_{ab}^*(x_ap_A,x_bp_B,p_t)\, H_{ab}(x_ap_A,x_bp_B,p_t)\, .
\nonumber\\
\label{eq:Fj-result}
\end{eqnarray}
We now rewrite the differences in denominations found above in Eq.\ (\ref{eq:Dij-diff-2}) as
\begin{equation}
D_i^{(j)}  - D_i^{(j+1)} = P^-\, \delta x^{(j)} + \hat \Delta_i^{(j)}  \, ,
\label{eq:Delta2}
\end{equation}
where 
\bea
\delta x^{(j)} \equiv \frac{1}{P^-}\, \sigma_j\, \left(\beta_{p_t}\cdot k_j - \beta_{Q-p_t}\cdot k_j \right)
\label{eq:Delta}
\eea
is independent of $i$. Now, using Eq.~(\ref{eq:Delta}), we rewrite Eq.~(\ref{eq:Fj-result})   as
\begin{eqnarray}
{\cal F}^{(j)} &=& \Bigg [\, \left(\prod _{i'=j+1}^S \frac{1}{D_{i'}^{(j+1)}+P^- \delta x^{(j)}+ \hat \Delta_{i'}^{(j)}  -i\epsilon}\right) 
\left(\prod _{i=1}^j \frac{1}{D_{i}^{(j+1)}+P^-\delta x^{(j)}+ \hat \Delta_i^{(j)} +i\epsilon } \right )-  
\nn\\
&\ & -\ \left(\prod _{i'=j+1}^S \frac{1}{D_{i'}^{(j+1)}-i\epsilon}\right)
 \left(\prod _{i=1}^j \frac{1}{D_{i}^{(j+1)}+i\epsilon}\right) \, \Bigg ]\ H_{ab}^*(x_ap_A,x_bp_B,p_t)\, H_{ab}(x_ap_A,x_bp_B,p_t)\, .
  \label{eq:Fj-result-2}
\end{eqnarray}
As at lowest order, we can neglect the dependence of the hard functions on soft momenta, so that they give an overall factor.  Then, as in Eq.\ (\ref{eq:AandD}), by a small shift, $\delta x^{(j)}$ in the collective partonic fraction $x$ applied to the first term in Eq.~(\ref{eq:Fj-result-2}) one can {\it simultaneously} absorb the differences between all 
denominators 
into the the smooth dependence of the remainder of the cross section, up to $i$-dependent corrections $\hat \Delta_i^{(j)}$, which are ${\cal O}(k^2)$  as discussed in Sec.\ \ref{subsec:loworder}. 
In turn, this implies that the terms inside the square brackets of Eqs.~(\ref{eq:Fj-result}) and (\ref{eq:Fj-result-2}) 
are suppressed by the ratio
\bea
\frac{\hat \Delta^{(j)}_i}{D_i^{(j+1)}} \sim \frac{k}{m_t}\, ,
\eea
where we have used that each denominator $D_i^{(j)}$ behaves linearly in soft momenta $k$ when expanded around the pinch surface.

In summary, as in the example of Section \ref{subsec:loworder},  the exchange of soft gluons in the final state does not contribute at leading power to 
 Eq.~(\ref{eq:result}), the single-particle inclusive cross section for the top quark in top pair production.

\subsection{Power corrections}
\label{ref:px}

Because all divergences are logarithmic, the suppression we have just found is enough to eliminate long-distance behavior to any fixed order in perturbation theory from an arbitrary pinch surface involving final state interactions.     Beyond fixed orders, however, soft gluons may be radiated by a spectator whose momentum is essentially nonperturbative.   Even though order-by-order the contributions from soft gluon exchange are small, higher-order corrections associated, for example, with the running of the QCD coupling and its infrared Landau  pole, suggest that the perturbative series for the infrared region of momentum space diverges.   Such reasoning, in fact, reconciles perturbation theory with the operator product expansion \cite{Mueller:1984vh}, and  underlies the treatment of power corrections in deep-inelastic scattering \cite{Dasgupta:1996hh} and event shapes in electron-positron annihilation \cite{Webber:1994cp}.       

For the hadronic scattering we are considering, this general reasoning suggests the presence of non-perturbative corrections associated with FSI.     To estimate the nature of the power corrections in this case, we will follow Ref.\  \cite{Qiu:1990xy} and interpret the finite remainder from the  region of soft gluon exchange as an additive nonperturbative correction to leading-power factorization.   To be specific, when the sum over final states produces a suppression of the form $\langle k \rangle/Q$ relative to leading-power behavior, with $\langle k\rangle\sim \Lambda \sim \Lambda_{\rm QCD}$ the size of the nonperturbative region and with $Q$ the hard scale, we infer that in the full theory there is an additive correction to the leading-power factorized cross section of size $\Lambda/Q$.  
Again, for heavy quark production, the exchanged gluon may originate from a spectator whose transverse momentum relative to the beam is at a nonperturbative scale set by $\Lambda_{\rm QCD}$.   In the case at hand, we thus infer that nonperturbative corrections in Eq.\ (\ref{eq:FTh})  for single-particle annihilation cross sections due to FSI are indeed an expansion in powers of $\Lambda/m_t$.    For practical considerations, such corrections are  presumably negligibly small, given the size of the hard scale.

\section{Multiparticle cross sections and corrections to factorization}
\label{sec:extensions}

As we have seen, the cancellation of final state interactions 
for a single-particle inclusive top cross section depends on combining contributions with different antitop momenta.   
This was built into our argument by `routing' the soft momentum differently for different choices of out state,
that is, by letting the antiquark recoil against final state momentum transfers due to the field of the spectators.   In particular, we combine
final states where $p_{\bar t}=Q-p_t$ and $p_{\bar t}=Q-p_t-k$.  This re-routing works so long
as the short-distance process that produces the top/antitop pair is insensitive to changes of
order $k$ in the antitop momentum.   
As we have just seen, corrections to this approximation are of 
order $k/m_t$ in the single-particle inclusive case.

We can generalize our arguments to multi-particle cross sections, starting with two-particle inclusive (2PI). We begin by noting that in a 2PI cross section we hold fixed $p_t$ {\it and} $p_{\bar t}$. It is then intuitively clear from our 1PI discussion that if we hold both the top and antitop momenta fixed we need some other particle or particles to recoil against the soft radiation in order to cancel final state
scatterings of both elements of the top pair. 
In the following we confirm
this assertion in two steps: first, we show that 
the cancellation mechanism fails
if there is no additional hard particle in the scattering process, and second, we 
identify
the pattern of cancellation of final state interactions when such additional radiation is present.
The same routing of soft momenta is necessary to fix
the somewhat more inclusive pair transverse momentum distribution,
so we expect the estimates of corrections that we derive below to apply in this case as well.

Consider a 2PI cross section with no additional hard radiation present (except the top and antitop), with Fig.\ \ref{fig:NLO-cuts} an example of such a configuration. Then, instead of routing the soft momentum $k$ through the antiquark to maintain the same $p_t$ for both out states
of the diagram, we would have to route the soft momentum through lines labelled $l$ and $l'$ in the figure,
in such a way that it appears in initial state light cone denominators like Eq.\ (\ref{initialDA}).    In these denominators, however
we cannot expand in $k$ unless it is much smaller than $\delta l$, which is in general at a hadronic scale,
precisely the scale of $k$, in fact.    In this case, the final state interactions fail to cancel, 
for much the same reasons as for the example given in Ref.\ \cite{Collins:2007nk} in a polarized
cross section.  
In our case, the effect is spin-independent and potentially leading power, making it impossible
to apply collinear factorization.    
In such a case, pair production without additional radiation, we expect leading power corrections from
final state interactions, as in the effects of rescattering found in Ref.\ \cite{Brodsky:1968rd} for
the photoproduction of electron-positron pairs on nuclei.
In the context of collider physics, restrictions on the phase space available for radiation
lead as well to logarithmic corrections at leading power.    For example, if the transverse momentum of 
radiation is limited to some value $\Lambda_{\rm QCD}\ll p_T\ll m_t$, then we will expect 
the cross section to be collinear factorizable up to corrections like $\Lambda_{\rm QCD}/p_T$.
At the same time, logarithms like $\ln(m_t/p_T)$ will appear, and for many observables
these logarithms have not yet been fully resummed or otherwise understood.
The most striking example is perhaps the ``superleading" logarithms \cite{Forshaw:2006fk},
which result from a partial non-cancellation that occurs
when an upper limit is put on soft radiation in a finite region of phase space.

If we do not limit additional radiation,
however, precisely the same arguments as in the previous section can be applied to show the cancellation
of final state interactions in 2PI cross sections for top and antitop, so long there is an additional
high-$p_T$
jet (taken relative to the collision axis, not the LC axis above)   
 as long as the cross section is inclusive enough in the momentum
of the jet so that shifts by momenta of order
$k$ lead to states that are counted.
In the following, we model the jet by a single parton of momentum $p_{\rm jet}$,
a simplification that does not weaken the argument.

We start by assuming that $p_{\rm jet}\cdot p_c \sim m_t^2$ for the other partons $c$ that take part in
the hard scattering:
$c=a,b,t,\bar t$. The production of the jet can be thought of as
local, and exactly the same argument holds  -- we simply
route the soft momentum $k$ through the high-$p_T$ jet rather than through the antitop.
We then compute
\bea
D_i^{(j+1)}-D_i^{(j)} &=& - [p^{(j+1)}_{t,i}]^-  - [p^{(j+1)}_{\bar t,i}]^ - - [p^{(j+1)}_{{\rm jet},\, i}]^-+ [p^{(j)}_{t,i}]^-  + [p^{(j)}_{\bar t,i}]^ -
 + [p^{(j)}_{{\rm jet},\, i}]^- \, ,
\label{eq:D-differences-pair}
\eea
where $p^{(j)}_{{\rm jet},\, i}$ is the momentum of the additional final state parton recoiling against 
the quark pair in state $i$.   As usual, the superscript $(j)$ labels the out state.    For this discussion,
soft momenta that flow through the top {\it and } antitop lines are routed in such a way
that the top and antitop momenta of the out state are fixed.

As above the differences can be compensated by a shift that is linear in the routing of
soft momenta, which depends on $j$ but which is the same for every $i$,
\begin{equation}
D_i^{(j)}  - D_i^{(j+1)} = P^-\, \delta x^{(j)} + \hat \Delta_i^{(j)}  \, .
\label{eq:Delta-pair}
\end{equation}
To derive $\delta x^{(j)}$, we recall the top momentum for each state $i$, and find in (\ref{eq:t-flow}), 
\begin{eqnarray}
[p_{t,i}^{(j+1)}]^- - [p_{t,i}^{(j)}]^- &=& -\  \sigma_j\, \beta_{p_t}\cdot k_j + {\cal O}(k^2) \, ,
\label{eq:pt-expand}
\end{eqnarray}
the same for each final state $i\ne j$.    We shall not need the explicit form of the
quadratic terms, but recall that they are an expansion in $k/m_t$,
which we consider negligible for this discussion.

Since we fix both the top and antitop 
momenta, we introduce a similar notation for the antitop's momentum, whose dependence on soft momenta is now
\begin{eqnarray}
[ p^{(j)}_{\bar t,i}\, ]^- &=& [\, p_{\bar t} - \sum_{l=i}^{j-1} \bar\sigma_l k'_l\, ]^-  ~~ {\rm for} ~~  i\le j\, ,  
\nonumber\\
&=& [\, p_{\bar t} + \sum_{l = j}^{i-1} \bar \sigma_l \bar k_l\, ]^-~~ {\rm for} ~~ i>j\, ,
\end{eqnarray}
where $\bar \sigma_l$ is defined by analogy to the top, positive for momentum
flowing in to the antitop line, and negative for momentum flowing out.   In this case, $p_{\bar t,j}^{(j)}=p_{\bar t}$,
the observed anti-top momentum.
Then we have again
\begin{eqnarray}
[p^{(j)}_{\bar t,i}]^- - [p^{(j)}_{\bar t,i}]^- &=& -\ \bar \sigma_j\, \beta_{p_{\bar t}} \cdot \bar  k_j + {\cal O}(k^2) \, ,
\label{eq:ptbar-expand}
\end{eqnarray}
independent of $i$, again neglecting terms explicitly suppressed by $k/m_t$.

The recoiling ``jet "line plays the role that was played by the antitop in the 1PI discussion, and
absorbs soft momenta of both the top and the antitop,
\begin{eqnarray}
[p^{(j)}_{{\rm jet},\, i}]^- &=& [\, \tilde p_{{\rm jet},\, i}  - \sum_{l\ge j} \sigma_l k_l -  \sum_{l\ge j} \bar \sigma_l \bar k_l\, ]^-  \, .
\label{eq:jet-recoil}
\end{eqnarray}
Here $\tilde p_{{\rm jet},\, i}$ is defined by analogy to $\tilde p_{\bar t, i}$
in Eq.\ (\ref{eq:tbar-recoil}), and is the momentum of the jet line in state $i$, not including soft
momenta routed through the jet line from the final state interactions of the 
top and antitop quark after the cut ({\it i.e.}, in the complex conjugate amplitude).
The difference between jet momenta with out states $j+1$ and $j$ is then given by the expansion
\bea
[p^{(j+1)}_{{\rm jet},\, i}]^- - [p^{(j)}_{{\rm jet},\, i}]^- 
&=& \beta_{\tilde p_{{\rm jet},\, i}}\, (\sigma_j k_j + \bar \sigma_j \bar k_j)
\nonumber\\
&\ & +\
\frac{2}{\tilde p_{{\rm jet},\, i}^+} (\sigma_j k_j + \bar \sigma_j \bar k_j) \cdot
\lrg_{\tilde p_{{\rm jet},\, i}} \cdot (\sum_{l\ge j} \sigma_l k_l + \sum_{l\ge j} \bar \sigma_l \bar k_l)
\nonumber\\
&\ & +\
\frac{1}{\tilde p_{{\rm jet},\, i}^+} (\sigma_j k_j + \bar\sigma_j \bar k_j) \cdot
\lrg_{\tilde p_{{\rm jet},\, i}} \cdot (\sigma_j k_j + \bar\sigma_j \bar k_j)
\, .
\label{eq:qi-expand2}
\eea
Combining this expression with (\ref{eq:pt-expand}), (\ref{eq:ptbar-expand}) 
to derive the difference of denominators in (\ref{eq:D-differences-pair}) and
then expanding, we find
\begin{eqnarray}
D_i^{(j+1)}-D_i^{(j)} &=&
-
\sigma_j \left( \beta_{\tilde p_{{\rm jet},\, i}} -\beta_{p_t}\right) \cdot k_j
- \bar \sigma_j \left( \beta_{\tilde p_{{\rm jet},\, i}} -\beta_{p_{\bar t}}\right)\cdot \bar k_j
 + \hat \Delta^{(j)}_i\, ,
\end{eqnarray}
where now $\hat\Delta^{(j)}_i$ is an expansion in $k/p_{\rm jet}^+$ as well as $k/m_t$.
Up to such quadratic terms we again find that an overall shift for each choice of $j$
leads to cancellation of final state interactions just as for the 1PI case.

Such high-$p_T$ jets, however, are absent in many, if not most, out states.
When the recoiling momentum, $p_{\rm jet}$ is much softer than the pair mass, 
or when it is far forward or backward, scales smaller than $m_t$
can come into play.
We can illustrate these corrections by considering the
general form of the hard scattering when it involves an extra final state gluon, of momentum $p'$.
In this case, the leading behavior is 
\bea
H_{ab\to t\bar t g(p')}^*(x_ap_A,x_bp_B,p_t,p_{\bar t},p')\, H_{ij\to t\bar t g(p')}(x_ap_A,x_bp_B,p_t,p_{\bar t},p')
&=& 
\nn\\
&\ & \hspace{-90mm} 
\sum_{c,d=a,b,t,\bar t} \frac{p_c \cdot p_d}{p_c\cdot p'\, p_d\cdot p'}H_{ij\to t\bar t}^*(x_ap_A,x_bp_B,p_t,p_{\bar t})\, H_{ij\to t\bar t}(x_ap_A,x_bp_B,p_t,p_{\bar t})\, ,
\eea
where to simplify this discussion of kinematic factors, we 
suppress color dependence.
In general, we would expect important contributions from $c,d = a,b$, the light partons that initiate the 
hard scattering, for which
\bea
\frac{p_a\cdot p_b}{p_a\cdot p'\, p_b\cdot p'} = \frac{1}{2}\, \frac{1}{p'_T\, {}^2}\, ,
\eea
in terms of the squared transverse momentum of the gluon (jet) relative to the beam axis.   
Corrections then arise because $p'$ depends on the 
choice of out state $j$.    Expansions of the hard scattering are
of the form
\bea
\frac{1}{(p' \pm k)^2_T} - \frac{1}{p'_T\, {}^2}\, ,
\eea
considered as a series in $k/p'_T$.   
For an azimuthally-symmetric cross section,
we expect corrections to go as even powers
of $1/p'_T$, but for spin-dependent cross sections,
for example, odd powers can arise.    
Following the reasoning of Sec.\ \ref{ref:px},
we infer that nonperturbative corrections to collinear factorization are
suppressed by powers of $\Lambda/p'_T$, in terms of some hadronic
scale $\Lambda$.

In summary, for two-particle inclusive and pair transverse momentum distributions,
 whenever the transverse momentum of recoiling radiation is much less than the pair mass,
we have seen that nonperturbative  corrections to collinear factorization 
are generally suppressed by powers of $\Lambda/p'_T$, where
$p'_T$ is the largest transverse momentum of an additional jet, rather than by powers of the top quark mass.

\section{Summary and discussion}
\label{sec:discussion}

\subsection{General considerations}

We have studied final state interactions (FSI) between hard partons produced in high-energy hadronic collisions and beam remnants from the initial state, and have shown that FSI cancel for a large class of inclusive cross sections at leading power,  to all orders in perturbation theory.  
Examples of processes for which these results are directly applicable include the forward-backward asymmetry of top-quark pairs at the Tevatron and a variety of processes with jets or identified hadrons (light or heavy quark fragmentation).

We have given our derivation in the language of light cone ordered perturbation theory.  While the nature of our proof is somewhat technical, the physics underlying our findings is quite transparent: cancellation of FSI occurs when additional, unobserved, hard radiation is allowed in the final state. The role of this additional radiation is to recoil against the observed system of hard partons and thus effectively absorb the kinematic effects of rescattering between the final state hard system and the beam remnants.
 
We find that the details of the cancellation depend on the nature of the final state. For top pair production and related processes, the size of  corrections due to interactions with spectators  are strongly suppressed for 1PI cross sections  \cite{Nayak:2005rt} but may be large for substantial portions of the total 2PI cross sections, when recoiling radiation is suppressed (because of jet vetoing, for example). Such corrections can appear both as higher-order perturbative corrections to a factorized cross section and as nonperturbative corrections suppressed by powers of perturbative scales.  The particular scales, however, depend on the set of final states.

For single particle inclusive observables, like single top in top pair production, we find that the scales suppressing the final state interactions are the large scales in the problem, like $p_t^+$ and $m_t$. The arguments we have given are valid as long as $p_t^+$ and $p_{\bar t}^+=(Q-p_t)^+$ 
are much larger than hadronic scales, including $\Lambda_{\rm QCD}$,
which is clearly the case for the top quark.
\footnote{Recall that $Q$, defined in (\ref{eq:Q-def2}), is the total initial state momentum available to the partonic hard scattering.}
In the massless limit, on the other hand, the validity of the expansions depend on the specific observable.  For $p_T$ distributions of massless partons, large transverse momenta are necessary to ensure a hard scattering. 
Then our considerations apply even in the massless case, so long as there is a hard particle recoiling against the observed particle.

We have found possible nonperturbative corrections to the leading power, factorized cross section of order $\Lambda_{\rm QCD}/q_T$, where again $q_T$ is the maximum transverse momentum of additional perturbative radiation in a given set of final states. Note that even for $q_T\gg \Lambda_{\rm QCD}$ a lower choice of $\mu_{\rm fact}\le q_T$    would also be necessary for the factorization scale in the leading-power term, rather than the hardest scale in the problem ($m_t$ in our case).   The mechanism of cancellation that we have identified, however, applies to all of these cases.
   
Our considerations generalize a pattern identified in Refs.\ \cite{Collins:2007nk,Rogers:2010dm}, showing how corrections associated with final-state interactions can become important in inclusive cross sections with two observed particles.   This is the case even as corrections to single-particle inclusive cross sections remain suppressed by the hardest scale of the final state \cite{Nayak:2005rt}.   Note that this means that calculations of charge asymmetries based on single-particle inclusive cross sections are stable against final-state interactions, but that limitations on the cross section, involving restrictions on quark pair momenta, or jet vetoes, can introduce sensitivity to nonperturbative scales.

We note that color and spin played no direct role in our analysis. The reason is that the FSI cancellation can be demonstrated by summing the cuts of one single (squared) diagram at a time. On the other hand, if one would like 
a computation or at least a better theoretical understanding of the remainder after the FSI cancellation, 
then theory-dependent group and spin factors will  of course
have to be taken into account. A toy model study of  color 
reconnection effects in hadronic final states is described in Refs.~\cite{Skands:2007zg,Wicke:2008iz} from a parton shower perspective.
For photoproduction of jets, a related analysis \cite{Luo:1994np} showed that the class of  power corrections
associated with soft gluon exchange in nuclei (collinear-) factorizes into higher-twist nuclear matrix elements, and are
in this sense universal.   For top pair production in hadronic collisions, we cannot anticipate such a higher-twist factorization,
precisely due to a mismatch in color factors associated with initial and final states,
as emphasized in \cite{Collins:2007nk,Rogers:2010dm}.

To close this brief discussion,
we note some possible directions for extending the present work:
\begin{enumerate}
\item A more detailed assessment of the remainders of 
the cancellation of FSI. Our arguments show that the remainder scales as the first power of $\langle l_T\rangle$. It is plausible, however, that 
in certain processes, the
first power nonperturbative correction vanishes. 
This appears to be the case, for example, for inclusive Drell-Yan production \cite{Beneke:1995pq,Laenen:2000ij}.
Another interesting possibility, suggested by analogy to the QED analysis in Low's theorem \cite{Low:1958sn,DelDuca:1990gz}, is that for top production the first nonleading 
power may be closely related to the derivative of the short-distance cross section.
More work is needed to 
explore
these possibilities. Clearly this is relevant since effects suppressed by the second power would likely not be experimentally 
accessible 
at high-energy hadron colliders while, as we discuss below, 
corrections suppressed by a single power of the hard scale
might well be observable.

\item Extension to other processes beyond top pair production. In this work we focused on top pair production because of its phenomenological relevance and simplified treatment of final state radiation. On the other hand, jet vetoes are very actively studied \cite{Marzani:2012wp,DuranDelgado:2011tp,Becher:2012qa,Banfi:2012jm,Berger:2010xi}, notably in Higgs boson production. The effect of the jet veto is to introduce logarithms of the ratio of the large hard scale and the presumably much smaller veto scale. It is for this reason that, at the few-percent level, FSI will overlap with the effects in resummations performed for jet vetoes, which motivates the need for a better understanding of FSI.

\item Extension to multi-scale kinematics. We often refer to large $p_T$ as the relevant 
hard scale in hadronic collisions, assuming we are reasonably inclusive in rapidities. However, in certain kinematic regions rapidities can be large and the right scale then will be a function of both $p_T$ and $y$. Understanding scale-setting in multi-scale problems is an important open problem at present, with forward dijet production being a notable example. Ref.~\cite{Hamilton:2012np} 
reports
recent work in this direction. We believe that the analysis presented here offers an additional perspective 
on the complexity of this problem.
\end{enumerate}

\subsection{Phenomenological implications}

Phenomenologically, the effect of FSI between hard partons and beam remnants 
may
be relevant for observables where additional radiation is suppressed, {\it i.e.} more exclusive observables. Let us consider an observed hard state $H$ that is accompanied by unobserved hard radiation. For example, $H$ can be a single top in top pair production or a color singlet state like an electroweak vector or Higgs boson. Then, the inclusive observable $H+X$ is not very sensitive to FSI, in the sense that subleading power corrections are suppressed by the largest hard scale in the problem. On the other hand, the exclusive contributions $H+nj,~ n=0,1,\dots$ separately could be quite sensitive to FSI since, as we 
have argued,
in such observables FSI are suppressed by the inverse veto scale, which is much lower.    Thus, corrections could be as large as ${\cal O}(1/20)\sim 5\%$ for a typical cut of around $20~ {\rm GeV}$
and a nonperturbative scale of 1 GeV. While at first it might seem surprising that FSI can have different impact within the same reaction, 
the effects of final state interactions in each exclusive channel 
can cancel in the fully inclusive cross section.

It is interesting to consider our findings in the light of 
available experimental data. In a recent study, the ATLAS collaboration \cite{ATLAS:2012al} measured the gap fraction $f$ in top pair events, which is the ratio of the cross-section subject to a veto $Q_0$ and the cross-section without a veto:
\begin{equation}
f(Q_0) = {\sigma(Q_0)\over \sigma} \leq 1 \, .
\end{equation}
The veto $Q_0$ can be as low as $20~{\rm GeV} \ll m_t \approx173~{\rm GeV}$ which is the ``typical" hard scale in top production. 
Ref.~\cite{ATLAS:2012al} compares the data with various fixed order calculations (of leading and next-to-leading order) interfaced to parton showers.  At lower $Q_0$ there are substantial uncertainties  in both the data and theory predictions.  Nevertheless, assuming an eventual decrease in systematic uncertainties, a full comparison between theory and data should take into account the possibility of, for example,
$1/Q_0$ corrections, in the light of our findings above. The analysis of measurements such as these may  
make possible direct access to final-state interactions in hard processes.

\begin{acknowledgments}
We thank Yang Bai, Stan Brodsky, Rouven Essig, Simone Marzani, Jianwei Qiu and Ted Rogers for discussions.  This work was supported by the National Science Foundation,  grants PHY-0653342 and PHY-0969739. A.M. thanks the C.N. Yang Institute for Theoretical Physics for hospitality.\end{acknowledgments}

\end{document}